\documentclass[twocolumn,preprintnumbers,amsmath,amssymb,floatfix,prl]{revtex4}
\usepackage{graphicx}
\usepackage{dcolumn}
\usepackage{bm}

\begin{document}

\title{Ferromagnetic Peierls insulator state in ${\bm A}$Mg$_4$Mn$_6$O$_{15}$ (${\bm A=}$ K, Rb, Cs)}

\author{T. Yamaguchi$^1$}
\author{K. Sugimoto$^2$}
\author{Y. Ohta$^1$}
\author{Y. Tanaka$^3$}
\author{H. Sato$^3$}
\affiliation{$^1$Department of Physics, Chiba University, Chiba 263-8522, Japan}
\affiliation{$^2$Center for Frontier Science, Chiba University, Chiba 263-8522, Japan}
\affiliation{$^3$Department of Physics, Chuo University, Tokyo 112-8551, Japan}

\date{\today}

\begin{abstract}
Using the density-functional-theory based electronic structure calculations, we study the electronic 
state of recently discovered mixed-valent manganese oxides $A$Mg$_4$Mn$_6$O$_{15}$ 
($A=$ K, Rb, Cs), which are fully spin-polarized ferromagnetic insulators with a cubic crystal 
structure.  We show that the system may be described as a three-dimensional arrangement of the 
one-dimensional chains of a $2p$ orbital of O and a $3d$ orbital of Mn running along the three 
axes of the cubic lattice.  We thereby argue that in the ground state the chains are fully spin polarized 
due to the double-exchange mechanism and are distorted by the Peierls mechanism to make the 
system insulating.  
\end{abstract}


\maketitle


Magnetism and electronic transport properties of materials are closely related to each other; 
e.g., insulating transition-metal oxides are typically antiferromagnetic, and ferromagnetism 
usually goes hand in hand with metallicity \cite{Khomskii1,Khomskii2}.  One of the rare 
exceptions to this rule is a hollandite chromate K$_2$Cr$_8$O$_{16}$ 
\cite{Hasegawa,Sakamaki,Toriyama,Nakao,Sugiyama,Takeda,Yamauchi}, where the double exchange mechanism 
\cite{Zener,Anderson,deGennes,Fazekas} induces a three-dimensional (3D) full spin polarization 
in the system below $T_c=180$ K, and then the metal-insulator (MI) transition follows in its 
fully spin-polarized quasi-one-dimensional (1D) conduction band by the Peierls mechanism 
at $T_{\rm MI}=95$ K without affecting its 3D ferromagnetism \cite{Nishimoto, Toriyama, Streltsov}.  
Thus, the uncommon ferromagnetic insulating state is realized in the ground state of this material.  

\begin{figure}[thb]
\begin{center}
\includegraphics[width=0.95\columnwidth]{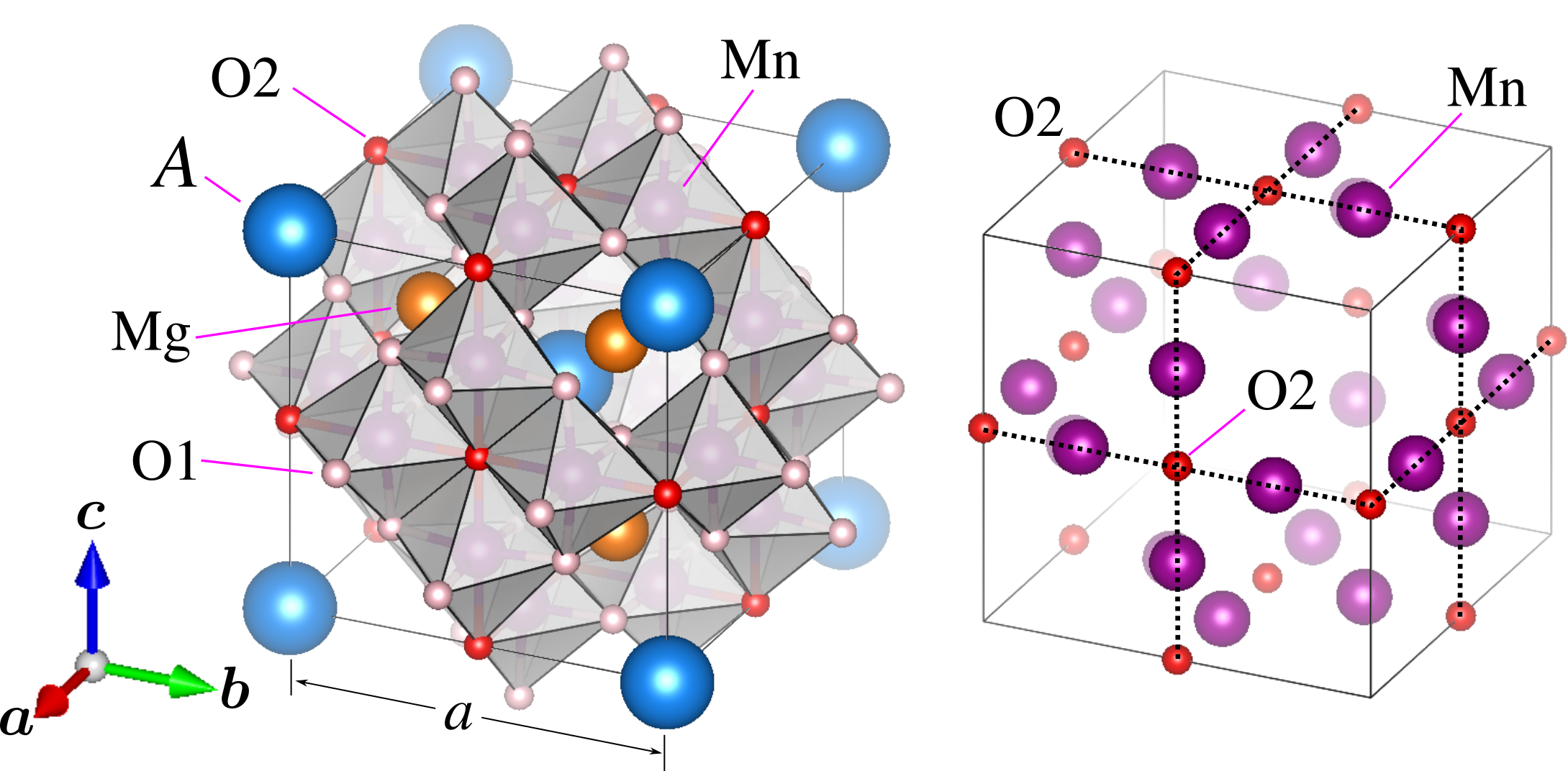}
\end{center}
\caption{Schematic representations of the crystal structure of $A$Mg$_4$Mn$_6$O$_{15}$ 
($A=$ K, Rb, Cs).  Atoms are distinguished by colors: $A$ (blue), Mg (orange), Mn (purple), 
O1 (pink), and O2 (red).  The 1D MnO chains are depicted in the right panel.  
}\label{fig1}
\end{figure}

Recently, Tanaka and Sato \cite{Tanaka} discovered a novel series of manganese oxides 
$A$Mg$_4$Mn$_6$O$_{15}$ ($A=$ K, Rb, and Cs), which were reported to be insulating 
ferromagnets with a highly-symmetric body-centered-cubic structure (see Fig.~\ref{fig1}).  
The Mn ions are in a mixed-valent state of Mn$^{3+}$ and Mn$^{4+}$ with an average oxidation 
state of $3.5+$ ($3d^{3.5}$) and are fully spin-polarized in the ground state with a ferromagnetic 
transition temperature of $T_c\simeq 170$ K.  The electric resistivity shows an insulating 
behavior in the entire temperature range observed (i.e., below 300 K).  
The materials reveal a large negative magnetoresistance: in KMg$_4$Mn$_6$O$_{15}$, 
the resistivity is suppressed by $\sim$ 40\% under 5 T of magnetic field.  

In this paper, we will show that a similar mechanism of insulating ferromagnetism to that of 
K$_2$Cr$_8$O$_{16}$ applies also in this manganese series.  Namely, we will use the 
density-functional-theory (DFT) based electronic structure calculations to demonstrate that 
an unexpectedly simple electronic state resides in this series with a rather complicated 
crystal structure: The ground state of the system may be described as a 3D arrangement 
of the three 1D chains of O $2p$ orbital and Mn $3d$ orbital, which are 
$p_\alpha-d_{3\alpha^2-r^2}-p_\alpha-d_{3\alpha^2-r^2}-\cdots$ ($\alpha=x,y,z$) 
running along the $\alpha$-axis of the cubic lattice.  
We will argue that the calculated localized/itinerant dualistic nature of electrons in the 
chains leads the system to ferromagnetism due to the double-exchange mechanism.  
We will also predict that these chains are dimerized by the Peierls mechanism, so that the 
system is insulating with a band gap in agreement with experiment; the system must be 
metallic if there were no lattice dimerizations.  
%



\begin{figure*}[tbh] 
\centering
\includegraphics[width=1.66\columnwidth]{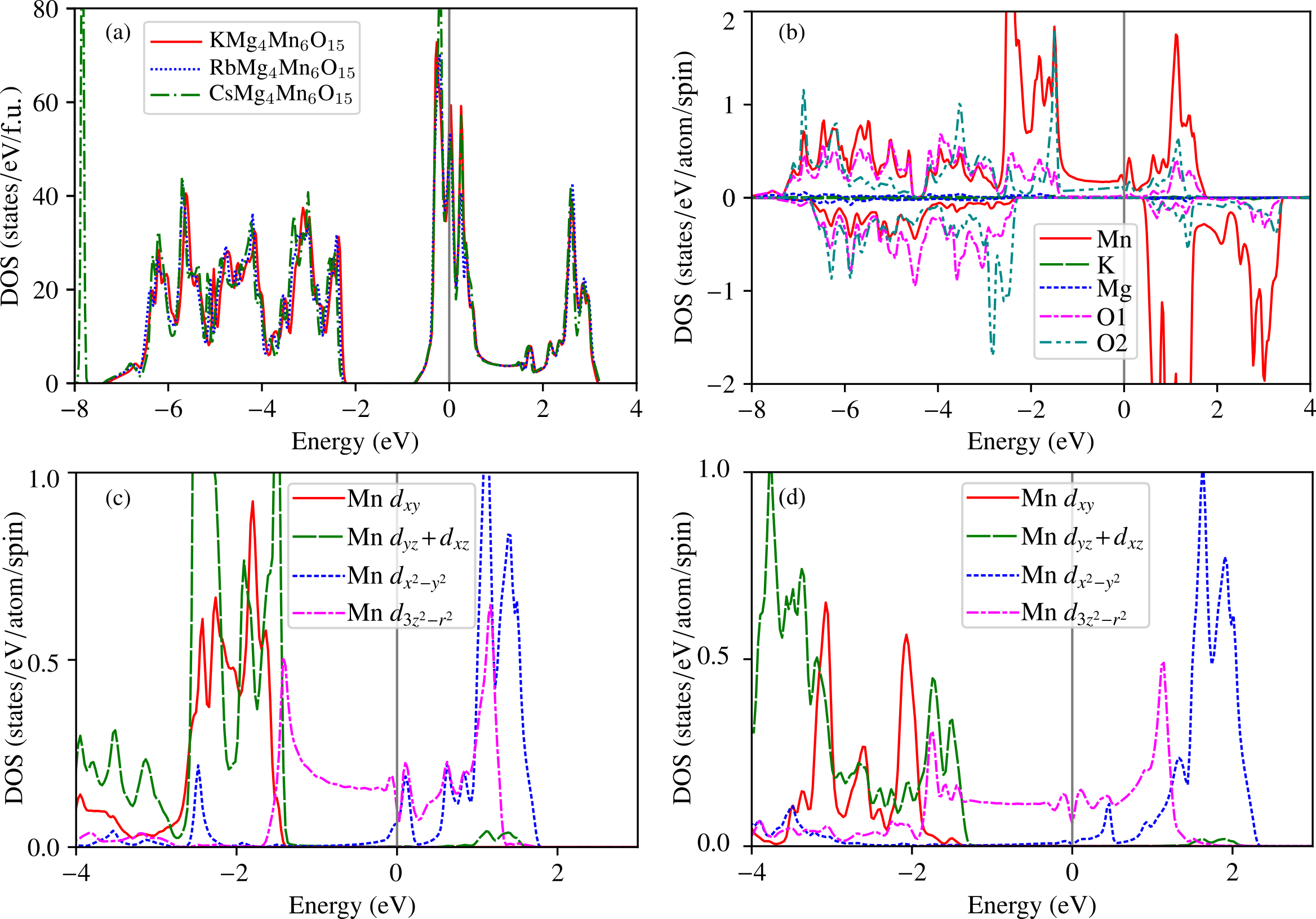}
\caption{Calculated total DOS and partial DOSs. 
(a) $A$-site dependence of DOS (per formula unit, f.u.) for the hypothetical paramagnetic 
$A$Mg$_4$Mn$_6$O$_{15}$ ($A=$ K, Rb, Cs) at $U=0$ eV.  A sharp peak at 
$\omega\simeq -8$ eV in $A=$ Cs comes from the $5p$ orbital of Cs ion.  
(b) Partial DOS projected onto each ion in ferromagnetic KMg$_4$Mn$_6$O$_{15}$ at $U=0$ eV, 
where the majority-spin (minority-spin) DOS is illustrated in the upper (lower) panel.  
A half-metallic situation is clearly seen.  
(c) Majority-spin partial DOS projected onto each $3d$ orbital of Mn in ferromagnetic 
KMg$_4$Mn$_6$O$_{15}$ at $U=0$ eV,  
(d) Same as in (c) but at $U=4$ eV.  
The vertical line in each panel represents the Fermi level.  
}\label{fig:dos}
\end{figure*} 


We employ the WIEN2k code \cite{Wien2k} based on the full-potential linearized 
augmented-plane-wave method for our DFT calculations.  We present calculated results 
obtained in the generalized gradient approximation (GGA) for electron correlations with 
the exchange-correlation potential of Ref.~[\onlinecite{perdew}].  To improve the description 
of electron correlations in Mn $3d$ orbitals, we use the rotationally invariant version 
of the GGA+$U$ method with the double-counting correction in the fully localized limit 
\cite{anisimov,liechtenstein}.  In the following, we will present the results obtained at 
$U=0$ and $4$ eV.  The spin polarization is allowed when necessary.  
The spin-orbit coupling is not taken into account in the following calculations, 
but we have checked that the spin-orbit coupling does not change our results 
qualitatively; e.g., the band gap does not open by the spin-orbit coupling.  

We use the crystal structure measured at room temperature \cite{Tanaka}, which has the 
cubic symmetry [space group $Im\bar{3}m$ (No.~229)] with the lattice constants of 
$a=8.3034(4)$, $8.3049(3)$, and $8.3476(5)$ for $A=$ K, Rb, and Cs, respectively, in units of \AA.  
The primitive unit cell contains 6 Mn and 15 O ions.  All the Mn ions are crystallographically 
equivalent but there are two crystallographically inequivalent O ions (which we call O1 
and O2).  There are 12 O1 and 3 O2 ions in the primitive unit call.  
In the self-consistent calculations, we use $15\times15\times15$ ${\bm k}$-points in the 
Brillouin zone.  Muffin-tin radii ($R_{\rm MT}$) of 2.50 ($A$), 1.96 (Mg), 1.94 (Mn), and 
1.67 (O) Bohr are used and we assume the plane-wave cutoff of $K_{\rm max}=8.50/R_{\rm MT}$.  
Because of a large $R_{\rm MT}$ value of $A$ ion, we choose the maximum value 
for the partial waves used in the computations of nonsphere matrix elements to be 6.  
We use VESTA \cite{momma} and XCrySDen \cite{kokalj} for graphical purposes.  



Now, let us discuss the calculated densities of states (DOSs), which are shown in Fig.~\ref{fig:dos}.  
First, we find in Fig.~\ref{fig:dos}(a) that the $A$-site dependence of DOSs is very small, 
in particular near the Fermi level, which is consistent with experiment where no qualitative 
differences in their electronic properties have been observed among $A$Mg$_4$Mn$_6$O$_{15}$ 
($A=$ K, Rb, and Cs) \cite{Tanaka}.  Hereafter, we will therefore discuss the electronic 
structure of KMg$_4$Mn$_6$O$_{15}$ only.  

Next, we show the calculated partial DOSs for KMg$_4$-Mn$_6$O$_{15}$ projected onto each ion 
in Fig.~\ref{fig:dos}(b), where the spin polarization is allowed and $U=0$ eV is assumed.  We find 
that the $3d$ orbitals of Mn are fully spin-polarized and form a half-metallic state, where the 
majority-spin band crosses the Fermi level but the minority-spin band exhibits a large band 
gap.  The calculated magnetic moment of $21\mu_{\rm B}$ per primitive unit cell is consistent 
with experiment \cite{Tanaka}.  The opposite spin polarization of $2p$ orbitals of O ion common 
in the negative charge-transfer-gap situation \cite{Sakamaki,Khomskii3,korotin} does not occur 
in the present case.  

Then, in Fig.~\ref{fig:dos}(c), we show the calculated majority-spin partial DOSs projected onto 
each $3d$ orbital of Mn ion at $U=0$ eV.  We find that the partial DOSs coming from the 
$t_{2g}$ orbitals are well localized around $-2$ eV, while those from the $e_g$ orbitals, the 
$d_{3z^2-r^2}$ orbital in particular, are rather extended between $-1.5$ and $1.5$ eV.  
This dualistic nature, i.e., the presence of both localized and itinerant electrons in the 
same system, suggests that the ferromagnetism of this system may be caused by the 
double exchange mechanism \cite{Zener,Anderson,deGennes,Fazekas}, just as in CrO$_2$ 
\cite{korotin,takeda} and K$_2$Cr$_8$O$_{16}$ \cite{Sakamaki,Toriyama}.  
The same situation also occurs at $U=4$ eV [see Fig.~\ref{fig:dos}(d)].  
We also note that the partial DOS curve of the Mn $3d_{3z^2-r^2}$ orbital exhibits 
the shape of DOS typical of the 1D tight-binding band, suggesting that the chain structure 
of Mn ions is formed in this system.  We will discuss this aspect further below.  

\begin{figure}[tb] 
\centering
\includegraphics[width=\columnwidth]{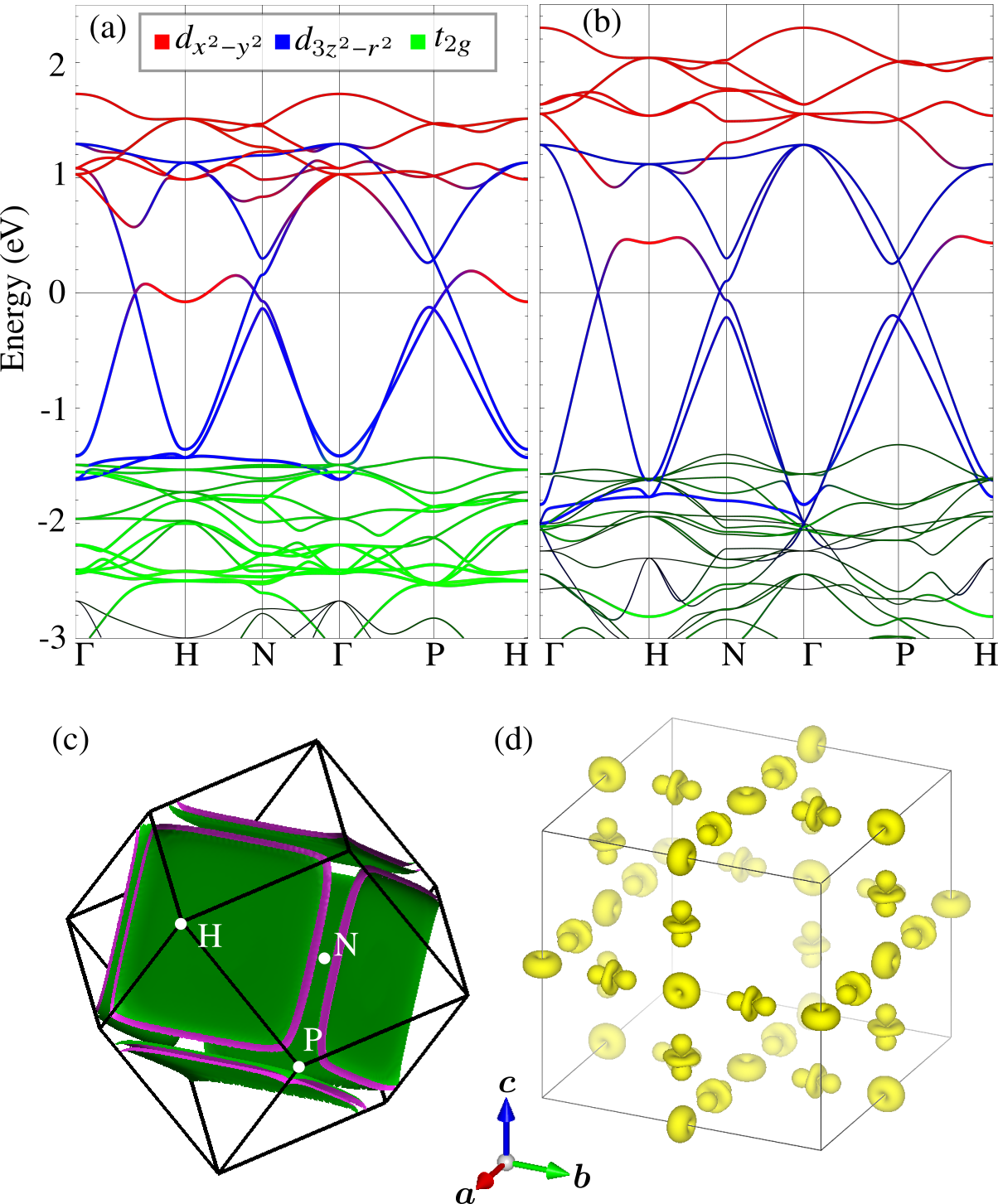}
\caption{
Calculated majority-spin band dispersions for the ferromagnetic phase of KMg$_4$Mn$_6$O$_{15}$ 
at (a) $U=0$ eV and (b) $U=4$ eV, where red, blue, and green curves represent the 
weight of the $d_{x^2-y^2}$, $d_{3z^2-r^2}$, and $t_{2g}$ contributions, respectively.  
The horizontal line in each panel indicates the Fermi level.  
(c) Calculated Fermi surfaces for the ferromagnetic phase of KMg$_4$Mn$_6$O$_{15}$ 
at $U=4$ eV.  The constant energy surfaces of 25 meV below the Fermi level are 
illustrated, so that the two sheets are slightly separated.  
(d) Calculated charge density distribution in the energy window of $\pm$ 0.1 eV around 
the Fermi level.  $U=4$ eV is assumed.  
}\label{fig:band}
\end{figure} 


The calculated majority-spin band dispersions of KMg$_4$Mn$_6$O$_{15}$ are shown in 
Figs.~\ref{fig:band}(a) and \ref{fig:band}(b) at $U=0$ and $U=4$ eV, respectively, where 6 red, 
6 blue, and 18 green curves representing the $d_{x^2-y^2}$, $d_{3z^2-r^2}$, and $t_{2g}$ contributions, 
respectively, are illustrated.  We find that at $U=0$ eV there is an electron pocket of the band 
coming predominantly from the $d_{x^2-y^2}$ orbital at the H point of the Brillouin zone.  This 
band shifts upward with increasing $U$, so that the electron pocket at the H point disappears 
at $U=4$ eV.  The bands forming the Fermi surfaces are thus predominantly from the 
$d_{3z^2-y^2}$ orbital.  

Note that the two bands cross each other at the Fermi level, giving rise to the ``surface-node'' 
Fermi surfaces, which are three pairs of the parallel flat plates, as shown in Fig.~\ref{fig:band}(c).  
The flat plates are made of two sheets with the ``nesting vector'' ${\bm Q}=0$, which indicates 
that when the unit cell contains more than two ions the Peierls instability causing the 
dimerization of ions may occur, keeping the unit cell unchanged.  A good nesting feature of the 
nesting vectors ${\bm Q}\simeq(2\pi/a,0,0)$, $(0,2\pi/a,0)$, and $(0,0,2\pi/a)$ is also noticed, 
indicating that the Peierls instability may also occur, which doubles the size of the unit cell of 
the system, i.e., from the body-centered-cubic structure to the simple-cubic structure (see below).  

To envisage the electronic state of the system in real space, we calculate the density 
distribution of electrons $\pm 0.1$ eV around the Fermi level.  The result is shown in 
Fig.~\ref{fig:band}(d), where we clearly find that the $3d_{3z^2-r^2}$ orbitals of Mn and one 
of the three $2p$ orbitals ($2p_z$) of O2 form the 1D chain structure along the $c$-direction 
of the cubic lattice.  Similarly, we find the chain structures formed by the $3d_{3x^2-r^2}$ orbital 
of Mn and $2p_x$ orbital of O2 along the $a$-direction and by the $3d_{3y^2-r^2}$ orbital of 
Mn and $2p_y$ orbital of O2 along the $b$-direction of the cubic lattce.  Note that the 
contributions from the O1 ions to the states near the Fermi level are very small.  

\begin{figure}[tb] 
\centering
\includegraphics[width=0.85\columnwidth]{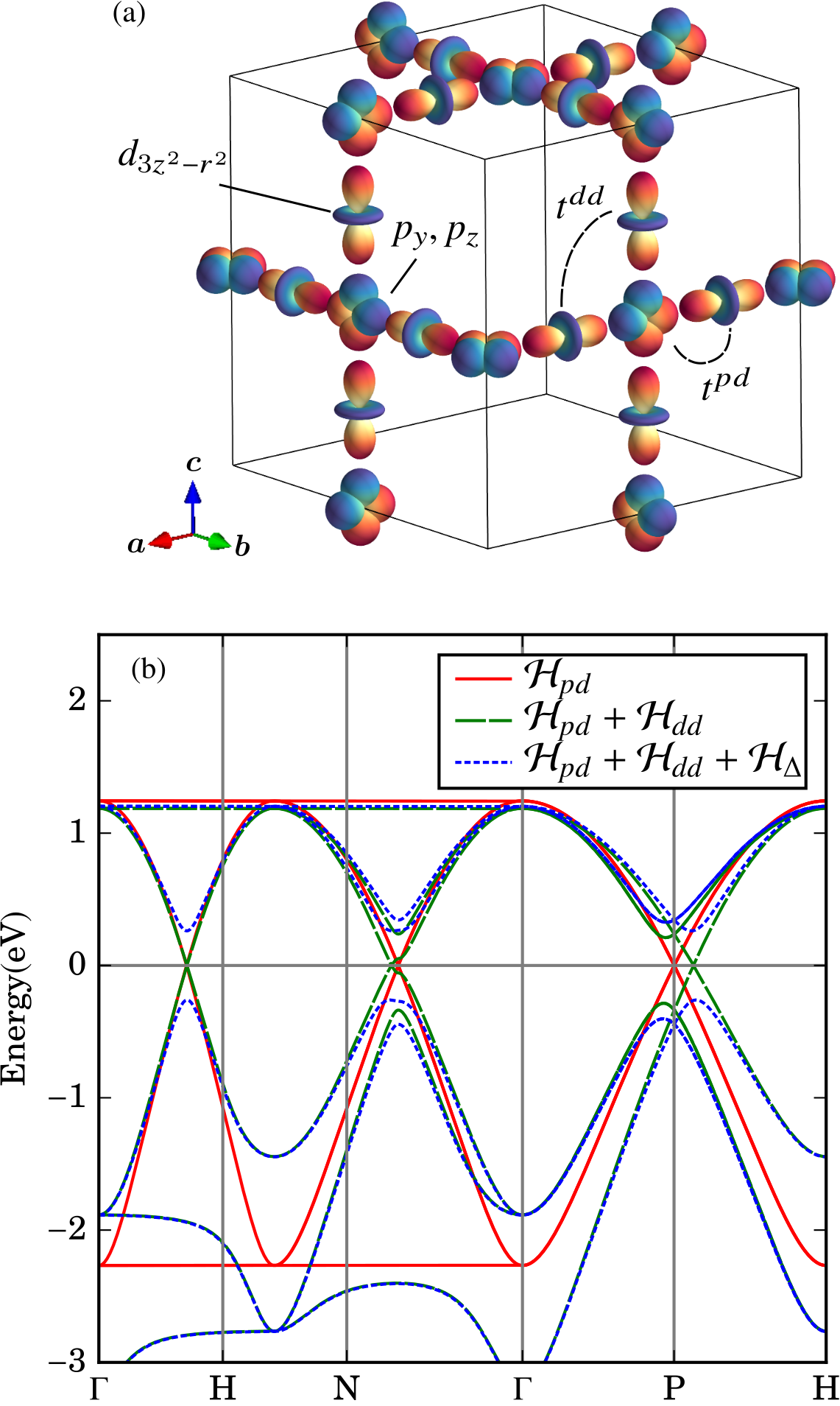}
\caption{
(a) Schematic representaion of the 1D chains of orbitals for KMg$_4$Mn$_6$O$_{15}$, where 
the orbitals $p_\alpha-d_{3\alpha^2-r^2}-p_\alpha-d_{3\alpha^2-r^2}-\cdots$ ($\alpha=x,y,z$) 
are illustrated.  
(b) Calculated majority-spin band dispersions of our tight-binding model, where we assume 
only $t^{pd}$ (red), both $t^{pd}$ and $t^{dd}$ (green), and $t^{pd}$, $t^{dd}$, and lattice 
dimerizations with alternating $\pm0.2$ eV modulations to $t^{pd}$ (blue).  
Six $d_{3\alpha^2-r^2}$ bands, some of which are degenerate, are drawn in each case.  
For the lattice dimerization, we assume the pattern illustrated in Fig.~\ref{fig:peierls}(a).  
}\label{fig:model}
\end{figure} 


Now, let us describe the low-energy electronic strucrture of this system by the tight-binding 
approximation, where the atomic orbitals form the 1D chains as shown in Fig.~\ref{fig:model}(a).  
The unit cell contains the 6 $3d$ orbitals ($d_{3z^2-r^2}$ and its equivalents) of Mn and 
6 $2p$ orbitals ($p_z$ and its equivalents).  The Hamiltonian for the majority-spin bands reads 
\begin{align*}
&\mathcal{H} = \varepsilon_{d}\sum_{i\mu} d^{\dagger}_{i\mu} d_{i\mu} 
  + \varepsilon_{p}\sum_{i\mu} p^{\dagger}_{i\mu} p_{i\mu} 
  + \mathcal{H}_{dp} + \mathcal{H}_{dd},  \label{TB} \\
&\mathcal{H}_{dp} = \sum_{\langle i\mu,j\nu\rangle} t^{pd}_{i\mu,j\nu} \big(d^{\dagger}_{i\mu} p_{j\nu} + \mathrm{H.c.}\big), \\
&\mathcal{H}_{dd} = \sum_{\langle i\mu,j\nu\rangle} t^{dd}_{i\mu,j\nu} \big(d^{\dagger}_{i\mu} d_{j\nu} + \mathrm{H.c.}\big), 
\end{align*}
where $d_{i\mu}^\dagger$ creates an electron on the orbital $\mu$ at Mn site $i$ and  
$p_{j\nu}^\dagger$ creates an electron on the orbital $\nu$ at O site $j$.  
$\langle i\mu,j\nu\rangle$ denotes the nearest-neighbor pair of orbital $\mu$ at site $i$ and orbital $\nu$ at site $j$.  
$\varepsilon_d$ and $\varepsilon_p$ are the on-site energies of Mn $3d$ and O $2p$ orbitals, respectively, 
and $t^{pd}$ and $t^{dd}$ are the hopping integrals between the neighboring $2p$ and $3d$ orbitals and 
between the neighboring $3d$ orbitals, respectively.  $\mathcal{H}_{dp}$ forms the 1D chains in the system 
and $\mathcal{H}_{dd}$ introduces the coupling between the chains giving rise to the 3D ferromagnetism.  
We calculate the maximally localized Wannier orbitals using the method of Refs.~[\onlinecite{mostofi,kunes}], 
which provides a good fitting of the band dispersions in a wide energy range with a large number of the 
tight-binding parameters.  However, we instead assume the values $\varepsilon_{p}=-4.5$, $\varepsilon_{d}=-2.5$, 
and $\left|t^{pd}\right|=2.2$ in units of eV and $t^{dd}/\left|t^{pd}\right|=-0.1$ for simplicity, which gives 
an accurate band dispersions at least near the Fermi level.  

\begin{figure}[tbh] 
\centering
\includegraphics[width=\columnwidth]{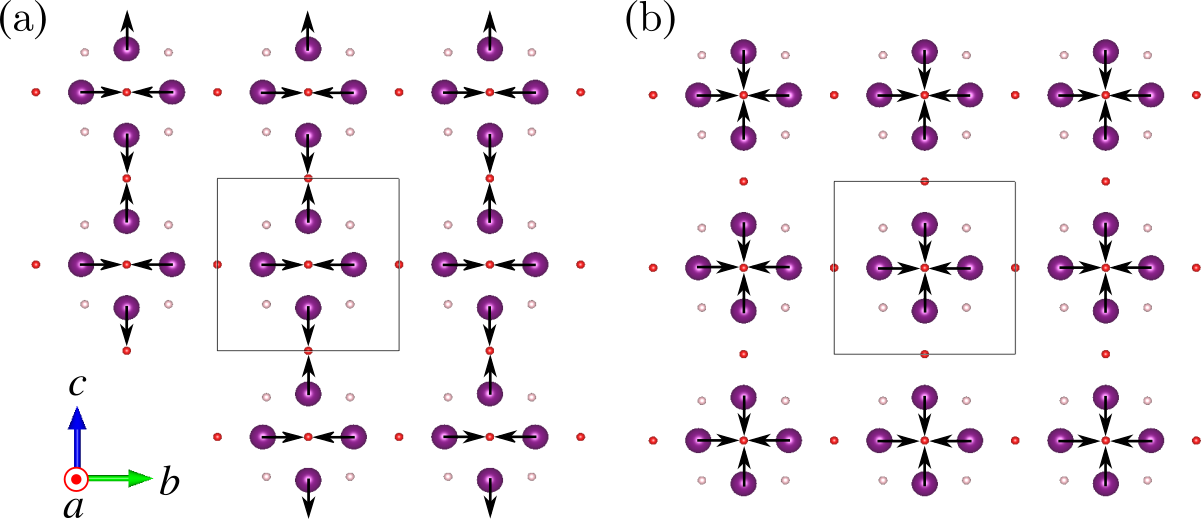}
\caption{Schematic representations of the lattice distortions 
(a) keeping the body-centered-cubic symmetry [space group $Im\overline{3}$ (No.~204)] and 
(b) keeping only the simple-cubic symmetry [space group $Pm\overline{3}m$ (No.~221)].  
The $[100]$ plane of the crystal, on which Mn ions are located, is illustrated.  
Arrows indicate the shifts of Mn ions along the 1D chain directions in the $[100]$ plane.  
}\label{fig:peierls}
\end{figure} 

The tight-binding bands thus obtained are shown in Fig.~\ref{fig:model}(b), where the results for three 
cases are plotted: 
(i) only $t^{pd}$ is included, 
(ii) both $t^{pd}$ and $t^{dd}$ are included, and 
(iii) $t^{pd}$, $t^{dd}$, and lattice dimerizations (adding a $\pm0.2$ eV alternation to $t^{pd}$, denoted 
as $\mathcal{H}_{\Delta}$) are included.  
We find that the inclusion of only $t^{pd}$ terms can reproduce the essential features of the bands, 
such as the crossing of the two bands at the Fermi level.  We also find that the addition of the 
$t^{dd}$ terms can explain the shift of the ${\bm k}$-point (from P to H) at which the two bands cross, 
as well as the lifting of the band degeneracy.  A better agreement with the results of the DFT-based 
band structure calculations in a wider energy range down to around $-2$ eV is obtained if we take 
into account the hopping integrals between Mn and O1 ions [see Fig.~\ref{fig:dos}(b)].  

Then, we find that the inclusion of the lattice dimerization actually leads to the opening of 
the band gap in the entire Brillouin zone, making the system insulating [see Fig.~\ref{fig:model}(b)].  
There are a variety of spatial patterns of the lattice dimerization (or relative phase of the Peierls 
distortions), but the pattern is unique if we assume that the primitive unit cell does not change, of 
which the pattern is illustrated in Fig.~\ref{fig:peierls}(a).  
If the primitive unit cell is extended (e.g., from the body-centered-cubic to simple-cubic lattices), 
we may have different patterns, of which an example is illustrated in Fig.~\ref{fig:peierls}(b).  
We made the DFT-based band structure calculations and checked that the band gap actually opens 
for the former pattern but the gap does not open for the latter, of which the results are found to be 
consistent with our tight-binding model calculations.  
We also made the structural optimization calculations based on the DFT, where we assume the space 
group $Im\overline{3}$ (No.~204) keeping the body-centered-cubic structure.  We thus obtained the 
structural distortion as shown in Fig.~\ref{fig:peierls}(a) and confirmed the opening of the band gap.  
We hope that further experimental studies will be made in future, to confirm 
the existence of the lattice distortion and to clarify what pattern is actually realized in the 
present materials.  

%
%
%

Finally, let us discuss the finite-temperature behavior of KMg$_4$Mn$_6$O$_{15}$ in comparison 
with that of K$_2$Cr$_8$O$_{16}$.  In the ground state, both materials are ferromagnetic insulators, 
where the double-exchange mechanism leads to ferromagnetism and the Peierls mechanism leads 
to the band gap formation.  We should however point out that, above the transition temperature 
of the ferromagnetic insulator state, K$_2$Cr$_8$O$_{16}$ is a ferromagnetic metal, while 
KMg$_4$Mn$_6$O$_{15}$ is a paramagnetic insulator.  The former situation is natural because we 
have a metallic band structure with the Peierls instability.  
However, the latter situation may also be possible if we consider as follows: 
The uniform magnetic susceptibility of KMg$_4$Mn$_6$O$_{15}$ obeys the Curie-Weiss law \cite{Tanaka}, 
indicating that the local magnetic moment persists even at high temperatures \cite{Moriya,Takahashi1,Takahashi2}.  
In other words, the ferromagnetic spin correlation extends to a spatially wide region even in the 
paramagnetic state at $T>T_c$, so that the fully spin-polarized electronic state remains locally and 
hence the Peierls mechanism of the lattice dimerization works there.  
In fact, a slow ferromagnetic spin fluctuation above $T_c$ has recently been observed by $\mu$SR 
experiment \cite{Okabe}.  
The spin fluctuation theory in the double-exchange ferromagnetism at finite temperatures should 
be developed in future to quantify this arguement.  


Summarizing, we have used the DFT-based electronic structure calculations to study the electronic 
state of recently discovered mixed-valent manganese oxides $A$Mg$_4$Mn$_6$O$_{15}$ ($A=$ K, Rb, Cs), 
which are fully spin-polarized ferromagnetic insulators with a cubic structure at the lowest 
temperatures.  
We have shown that the system may be described as a 3D arrangement of the 1D chains of a 
$2p$ orbital of O and a $3d$ orbital of Mn running along the three axes of the cubic lattice.  
We have argued that in the ground state the chains are fully spin polarized due to the 
double-exchange mechanism and are distorted by the Peierls mechanism to make the system 
insulating.  We have thus predicted the presence of the lattice dimerization in the wide temperature 
range and possible occurence of the Peierls metal-insulator transition at a much higher temperature, 
for which further experimental studies are desirable.  

We thank T.~Konishi, H.~Okabe, T.~Toriyama, and T.~Yamauchi for useful discussions.  
This work was supported in part by a Grant-in-Aid for Scientific Research 
(No.~17K05530) from JSPS of Japan.

\end{document}